\documentstyle[sprocl]{article}

\newcommand{\eq}{\begin{equation}}
\newcommand{\en}{\end{equation}}
\newcommand{\eqa}{\begin{eqnarray}}
\newcommand{\ena}{\end{eqnarray}}

\def\be{\begin{equation}}
\def\ee{\end{equation}}
\def\bea{\begin{eqnarray}}
\def\eea{\end{eqnarray}}

\begin{document}

\title{FINDING REGULATORY SITES FROM STATISTICAL
ANALYSIS OF NUCLEOTIDE FREQUENCIES IN THE UPSTREAM REGION OF
 EUKARYOTIC GENES.}
\author{M. Caselle$^a$ and P. Provero$^{a,b}$}
\address{$^a$ Dipartimento di Fisica Teorica, Universit\`a di Torino, and 
INFN,\\
  sezione di Torino, Via P. Giuria 1, I-10125 Torino, Italy.\\
  e-mail:~~ caselle@to.infn.it,~~~~  provero@to.infn.it}
\address{$^b$ Dipartimento di Scienze e Tecnologie Avanzate, Universit\`a del\\
Piemonte Orientale,
 I-15100 Alessandria, Italy.}

\author{F. Di Cunto and M. Pellegrino} 
\address{ Dipartimento di Genetica, Biologia e Biochimica, 
 Universit\`a di Torino,\\
 Via Santena 5 bis, I-10100, Torino, Italy.\\
e-mail: ferdinando.dicunto@unito.it}

\maketitle\abstracts{We discuss two new approaches to extract relevant
biological information on the Transcription Factors (and in particular to
identify their
binding sequences) from the statistical distribution of oligonucleotides in the
upstream region of the genes. Both the methods are based on the
notion of a ``regulatory network'' responsible for the various expression
patterns of the genes. In particular we concentrate on families of coregulated
genes and look for the simultaneous presence in the upstream regions of these
genes of the same set of transcription factor binding sites. 
We discuss two instances which
well exemplify the features of the two methods: the coregulation of 
glycolysis in {\it Drosophila melanogaster} and 
the diauxic shift in {\it Saccharomyces cerevisiae}.}

\section{Introduction}
\label{s1}
As more and more complete genomic sequences  are being decoded it is becoming
 of crucial importance
to understand how the gene expression is regulated.
A central role in our present understanding of gene expression is played by the
notion of ``regulatory network''. It is by now clear that a particular
expression pattern in the cell is the result of an intricate network of
interactions among genes and proteins 
 which cooperate to enhance (or depress) the 
expression rate of the various genes. It is thus important to address the
problem of gene expression at the level of the whole regulatory network and not
at the level of the single 
gene\cite{pg97,Wagner:1997,Tavazoie:1999,Pilpel:2001,Bussemaker:2001}.

In particular, most of the available information about such interactions
concerns the transcriptional regulation of protein coding genes.
Even if this is not the only regulatory mechanism of gene expression in
 eukaryotes it is certainly the most widespread one.

In these last years, thanks to the impressive progress in the DNA array
technology several results on these regulatory networks have been obtained.
Various transcription factors (TF's in the following) have been identified
and their binding motifs in the DNA chain (see below for a discussion) have been
characterized. However it is clear that we are only at the very beginning of
such a program 
and that much more work still has to be done in order to reach a satisfactory
understanding of the regulatory network in eukaryotes (the situation is somehow
better for the prokaryotes whose regulatory network is much simpler).

In this contribution we want to discuss a new method which allows to
reconstruct these interactions by comparing existing biological 
information with the statistical properties of the
sequence data.  This is a line of research which has been pursued in the last
few years, with remarkable results, by several groups in the world. For a
(unfortunately largely incomplete) list of references see 
\cite{Wagner:1997,Tavazoie:1999,Pilpel:2001,Bussemaker:2001,vanHelden:1998,vanHelden:2000,Hughes:2000,hw2000}.
 In particular the biological input that we shall use is the 
fact that some genes, being involved in the same biological process, are likely
to be ``coregulated'' i.e. they should show the same expression pattern.
The simplest way for this to happen is that they are all regulated by the
same set of TF's. 
If this is the case we should find in the 
upstream\footnote{With this term 
 we denote the portion of the DNA chain
which is immediately before the starting point of the open reading frame (ORF).
We shall characterize this region more precisely in sect.\ref{s3} below.}
 region of these
genes the same TF 
binding sequences. This is a highly non trivial occurrence from a
statistical point of view and could in principle be recognized by simple
statistical analysis.

As a matter of fact the situation is much more complex than what
 appears from this idealized picture.
 TF's not necessarily bind only to the upstream 
region. They often recognize more than one sequence (even if there is usually
a ``core'' sequence which is highly conserved). Coregulation could be achieved
by a complex interaction of several TF's instead than following the simple
pattern suggested above.
Notwithstanding this, we think that it is worthwhile to explore this
simplified picture of coregulation, for at least three reasons.
\begin{itemize}
\item
Even if in this way we  only find a subset of the TF's involved in the
coregulation, this would be all the same an important piece of information: 
It would add a new link in the regulatory network that we are studying.
\item
Analyses based on this picture, being very simple,
 can be easily performed on any gene set, from the few genes
involved in the Glycolysis (the first example that we shall discuss below) up to
 the whole genome (this will be the case of the second example that we shall
 discuss). This feature is going to be more and more important as more and more 
 DNA array experiment appear in the literature.  As the quantity of available  
  data  increases, so does the need  of analytical tools to analyze it.
\item
Such analyses could  be easily improved to include some of the features 
outlined above, keeping into account, say, the sequence variability or the 
synergic interaction of different TF's.
\end{itemize}

To this end we have developed two different (and complementary) approaches.
The first one (which we shall discuss in detail in sect.\ref{s3} 
below) follows a more 
traditional line of reasoning: we start from a set of genes which are known to
be coregulated (this is our ``biological input'') and then try to recognize the
possible binding sites for the TF's. We call this approach the ``direct
search'' for coregulating TF's. 

The second approach
(which we shall briefly sketch in sect.\ref{s4} 
below and is discussed in full detail
in~\cite{ccp}) is completely different and
is particularly suitable for the study of genome-wide DNA array experiments. In
this case the biological input is taken into account only at the end of the
analysis. We start by organizing all the genes in
 sets on the basis of the overrepresented common sequences and then filter them
 with expression patterns of some DNA array experiment.
We call this second approach the ``inverse
search'' for coregulating TF's.

It is clear that 
all the candidate gene interactions which we  identify
with our two
methods have to be tested experimentally.
However our results may help selecting among the huge number of possible
candidates and could be used as a preliminary test to guide the experiments. 

This contribution is organized as follows. In sect.\ref{s2}
 we shall briefly introduce
the reader to the main features of the regulatory network (this introduction
 will
necessarily be very short, the interested reader can find a thorough
discussion for instance in~\cite{book}).
 We shall then
devote sect.\ref{s3} and \ref{s4}
 to explain our ``direct'' and ``inverse'' search methods
respectively.
 Then we shall discuss two instances which well exemplify the two strategies.
First in sect.\ref{s5} we shall study the coregulation of 
glycolysis in {\it Drosophila melanogaster}. 
Second, in sect.\ref{s6} 
we shall discuss 
the diauxic shift in {\it Saccharomyces cerevisiae}.
The last section will be devoted to some
concluding remarks.

\section{Transcription factors.}
\label{s2}
As mentioned in the introduction, a major role in the regulatory network is
played by the Transcription Factors, which may have  in general  
 a twofold action on gene transcription.
 They can activate it by
  recruiting the transcription machinery to the transcription starting site
 by binding enhancer sequences in the upstream noncoding region, or by
 modifying chromatine structure, 
  but they can also repress it by negatively
  interfering with the transcriptional control mechanisms.

The main point is that 
{\sl in both cases TFs act by binding to specific, often short 
DNA sequences in the
upstream noncoding region}. It is exactly this feature which allows TF's to
perform a {\bf specific} regulatory functions.
These binding sequences can be considered somehow as the fingerprints of the
various TF's. The main goal of our statistical analysis will be the
identification and characterization of such binding sites.

\subsection{Classification}
\label{s2.1}
Even if TF's show a wide variability it is possible to try a (very rough) 
classification. Let us see it in some more detail, since it will help
understanding  the examples which we shall discuss in the following sections.
There are four main classes of binding sites in eukaryotes.

\begin{itemize}
\item  {\bf Promoters}

These are localized in the region immediately upstream of the coding
region (often  
within {200 bp} 
from the transcription starting point). They can be of two types:
\begin{itemize}
\item short sequences like the well known {CCAAT-box, TATA-box, GC-box}
which are {\bf not} tissue specific and are recognized by ubiquitous
TFs
\item tissue specific sequences which are only recognized by tissue specific TFs
\end{itemize}

\item  {\bf Response Elements}

These appear only in those genes whose expression is controlled by an external
factor (like hormones or growing factors). These are usually within {1kb}
 from
the transcription starting point. Binding of a response element
 with the appropriate factor
may induce a relevant enhancement in the expression of the corresponding gene  
\item  {\bf Enhancers}
 
these are regulatory elements which, differently from the promoters, can act in
both orientations and (to a large extent) at any distance from the transcription
starting point (there are examples of enhancers located even {50-60kb}
 upstream).
 They enhance the expression of the corresponding gene.

\item  {\bf Silencers}

Same as the enhancers, but their effect is to repress the expression of the
gene.
\end{itemize}

\subsection{Combinatorial regulation.}
\label{s2.2}
The main feature of TF's activity is its ``combinatorial'' nature. This 
means that:
\begin{itemize}
\item
 a single gene is usually regulated by many independent TF's which  bind to
 sites which may be very far from each other in the upstream region.
\item
it often happens that several TF's must be {\bf simultaneously} present
 in order to perform their regulatory function.
This phenomenon is usually referred to  as the ``Recruitment model for gene
activation'' (for a review see \cite{pg97})
 and represents the common pattern of action 
of the TF's. It is so important that it has been recently adopted as guiding
principle for various computer based approaches to detect regulatory sites (see
for instance \cite{Pilpel:2001}).
\item
the regulatory activity of a particular TF is enhanced if it can bind to 
several (instead of only one) binding sites  in the upstream region. This
``overrepresentation'' of a given binding sequence is also used in
some algorithms
which aim to identify TF's. It will also play a major role in our approach.
\end{itemize}

\section{The ``direct'' search method.}
\label{s3}
In this case the starting point is the selection of a set of genes which are
known to be involved in the same biological process (see example of sect.
\ref{s5}).

Let us start by fixing a few notations:

\begin{itemize}
\item
Let us denote with $M$ the number of genes in the coregulated set
and with $g_i$, $i=1\cdots M$ the genes belonging to the set
\item
Let us denote with $L$ the number of base pairs (bp) of the 
 upstream non coding region on which we shall perform our analysis.
It is important to define precisely what we mean by ``upstream region''
With this term 
 we denote the {\sl non coding} portion of the DNA chain
which is immediately before the transcription start site.
This means that we do {\bf not} consider as part of this region the UTR5 part of
the ORF of the gene in which we are interested. If we choose $L$ large enough it
may happen
that  other ORFs are present in the upstream region. In this case we consider as
upstream region only the non coding part of the DNA chain 
up to the nearest ORF  (even if it appears in the opposite strand).
Thus $L$ should be thought of as an upper cutoff. In most cases the length
of the upstream region is much smaller and is gene dependent. We shall denote it
in the following as $L(g)$.
\item
In this upstream region we shall be interested in studying short
 sequences of nucleotides which we shall call {\sl words}. Let $n$ be
the length of such a word. For each value of $n$ we have 
$N\equiv 4^n$ possible  words $w_i$, $i=1\cdots N$.
 The optimal choice of $n$ (i.e. the one which 
optimize the statistical significance of our analysis) is a function of $L$ and
$M$. We shall see some typical values in the example of sect.\ref{s5}
In the following we shall have to deal with words of varying size. When needed,
in order to avoid confusion, we shall call $k$-word a word made of $k$
nucleotides.

\end{itemize}
Let us call $U$
 the collection of upstream regions of the 
$M$ genes $g_1,\dots g_M$.
Our goal is to see if the number of occurrences of a
given word $w_i$ in each of the upstream regions belonging to $U$
 shows a ``statistically significant''  deviation
(to be better defined below) from 
what expected on the basis of pure chance.
To this end we perform two types of analyses.
\vskip0.5cm\noindent
{\bf First level of analysis.}
\vskip0.5cm\noindent
This first type of analysis is organized in three steps
\begin{itemize}
\item 
{\bf Construction of the ``Reference samples''}.  The first step is
the construction of  
 a set of $p$ ``reference samples'' which we call $R_i, 
i=1,\cdots p$.
The $R_i$ are {\sl nonoverlapping} sequences of $L_R$ nucleotides each, 
 extracted from a
noncoding portion of the DNA sequence 
 in the same region of the genome to which the genes that we study belong 
 but ``far'' from any ORF. 
 From these reference samples we then extract for each
 word the ``background occurrence probability'' that we shall then use as input
 of the second step of our analysis. The rationale behind this approach is the
 idea that the coding and regulating parts of the genome are immersed in a large
 background sea of ``silent'' DNA and that we may recognize that a portion of
 DNA has a biological function by looking at statistical deviations in the word
 occurrences with respect to the background. However it is clear that this is a
 rather crude description of the genome, in particular there are some obvious
 objections to this approach:
\begin{itemize}
\item
 There is no clear notion of what ``far'' means. As we mentioned in the
introduction one can sometimes find TF's which keep their regulatory function
even if they bind to sites which are as far as $\sim 50 kb$ from the ORF  
\item 
It is possible that in the reference samples the
 nucleotide frequencies reflect some unknown biological function thus inducing a
 bias in the results
\item 
It is not clear how 
should one deal with the long repeated sequences which very often appear in
the genome of eukaryotes
\end{itemize}
We shall discuss below how to overcome these objections.
\item
{\bf Background probabilities.} For each word $w$ we study the number
of occurrences $n(w,i)$ in the $i^{th}$ 
sample. They will follow a Poisson distribution from which we extract the
background occurrence probability of the word. This method works only if 
$p$ and $L_R$ are large enough with respect to the number of possible
words $N$ (we
shall see in the example below some typical values for $p$ and $L_R$).
However we have checked that our results are robust with respect to different
choices of these background probabilities.
\item 
{\bf Significant words.} From these probabilities we  can immediately
construct for each $n$-word the 
expected number of occurrences in each of the upstream sequences of
$U$ and from 
them the probabilities $p(n,s)$ of finding at least one $n$-word simultaneously
present in the upstream region of $s$ (out of the
$M$) genes. By suitably tuning $L$, $s$ and $n$
we may reach very low probabilities.
If notwithstanding such a low probability we indeed find a 
$n$-word which appears in
the upstream region of $s$ genes then we consider this fact as a strong
indication of its role as binding sequence for a TF's. We may use the
probability $p(n,s)$
as an estimate of the significance of such a candidate binding
sequence.
\end{itemize}

As we have seen the critical point of this analysis is in the choice of the
reference sample. We try to avoid the bias induced by this choice 
 by crossing the above procedure with a second
level of analysis 
\vskip0.5cm\noindent
{\bf Second level of analysis.}
\vskip0.5cm\noindent
The main change with respect to the previous analysis is that in this case we
extract the reference probabilities for the $n$-words
from an {\sl artificial} reference sample 
constructed with a Markov chain algorithm based on the  frequencies
of $k$-words with $k<<n$ (usually $k=1,2$ or 3) extracted from
the upstream regions themselves. Then the second and third step of the previous
analysis  follow unchanged. 
The rationale behind this second approach is that
we want to see if in the upstream region there are some $n$-words
(with $n=7$ or $8$, say) that occur much more often than
what one would expect based on the frequency of the $k$-words in the
same region.  

These two  levels of analysis are both likely to give
results that are biased according to the different choices of reference
probabilities that define them. However, since these biases are
likely to be very different from each other, it is reasonable to expect that by
comparing the results of the two methods one can minimize the number
of false positives found.   
\section{The ``inverse'' search method.}
\label{s4}
A major drawback of the analysis discussed in the previous section is that it
requires a precise knowledge of the function of the genes examined. 
As a matter of fact a large part
of the genes of eukaryotes have no precisely 
know biological function and could not be studied with our direct method.
Moreover in these last years the richest source of biological
information on gene 
expression comes form microarray experiments, thus it would be important to
have a tool to study gene coregulation starting from the output of such
experiments. These two observations suggested us the inverse search method that
we shall briefly discuss in this section. We shall outline here only the main
ideas of the method, a detailed account can be found in~\cite{ccp}.

The method we propose has two main steps: first the  ORFs 
of an eukaryote genome are grouped in
(overlapping) sets based on words  that are
overrepresented 
in their upstream region, with respect to their frequencies in a
reference sample which is 
made of all the 
upstream regions of the whole genome. 
Each set is labelled by a word.
Then for each of these sets the average
expression in one ore more microarray experiments are compared to the
genome-wide average: if a statistically significant difference is
found, the word that labels the set is a candidate regulatory site
for the genes in the set, either enhancing or inhibiting their
expression. 

An important  feature is that  the grouping of the genes into sets depends 
only on the upstream sequences and not on the microarray experiment
considered: It needs to be done only once for each organism, 
and can then be used to analyse an arbitrary number of microarray
experiments. 

We refer to \cite{ccp} for a detailed description of how the sets are
constructed, we only stress here that this construction only requires three
external parameters which must be fixed by the user: the length $L$ of the
upstream region (see sect.\ref{s3} for a discussion of this parameter), the
length $n$ of the words that we use to group the sets and a cutoff probability
$P$ which quantifies the notion of ``overrepresentation'' mentioned above.

\section{Example: glycolysis in Drosophila melanogaster.}
\label{s5}

As an example of the analysis of sect.\ref{s3}, we studied the 7 genes 
of {\it Drosophila melanogaster} involved
in glycolysis. These genes are listed in 
Tab.1.
\begin{table}[t]
\caption{Genes involved in the glycolysis.
\label{tab:one}}
\vspace{0.2cm}
\begin{center}
\begin{tabular}{|l|l|l|c|} \hline
Gene & Description &Locus & Chromosome \\ \hline\hline
Ald & Aldolase& AE003755 & 3R \\ \hline
Eno & Enolase& AE003585 & 2L \\ \hline
Gapdh1 &Glyceraldehyde 3-ph. dehydrogenase 1 & AE003839 & 2R \\ \hline
Gapdh2 & Glyceraldehyde 3-ph. dehydrogenase 2 & AE003500 & X \\ \hline
Hex &Hexokinase & AE003756 & 3R \\ \hline
ImpL3 &L-lactate dehydrogenase& AE003563 & 3L \\ \hline
Pfk & 6-phosphofructokinase& AE003755 & 2R \\ \hline
\end{tabular}
\end{center}
\end{table}
We performed our analysis with two choices of the parameters:
\begin{description}
\item{1]}  {\bf Promoter region.}
In this first test we decided to concentrate in the promoter region. Thus we
 chose $L\leq 100$. With this choice, and since $M=7$, we are bound to study
 $n$-words with $n=3,4,5$ in order to have a reasonable statistical
 significance. In particular we concentrate on $n=3$
In the first level of analysis we chose $L_R=100$ and $p=1000$ 
($p$ is the number of reference samples). In the second level 
of analysis we
chose $k=1$ ($k$ being the number of nucleotides of the $k$-words 
used to construct the Markov chain).
 We found (among few other motifs which we do not discuss here for
brevity) that a statistically relevant signal is reached by
the sequence {GAG}. 
This result has a clear biological interpretation
since it is the binding site of an ubiquitous 
 TF known as GAGA factor which belongs to the
 class of the so called ``zinc finger'' TF's\footnote{The commonly assumed
 binding site for
 the GAGA factor is the sequence GAGAG, however it has been recently realized
 that the minimal binding sequence is actually the 3-word GAG~\cite{gag}.}.
We consider this finding as a good validation test of the whole procedure.

\item {2]} {\bf large scale analysis}

In this second test we chose $L=5000$. This allowed us to address 
$n$-words with
$n=6,7,8$. For the reference samples we used $L_R=5000$ and $p=21$
As a result of our analysis we obtained the probabilities $p(n,s)$ of finding at
least one $n$-word in the upstream region of
 $s$ out of the 7 genes that we are studying. As an example we list in Tab.2 the
 values of $p(n,s)$ for $s=7$ and $n=6,7,8$. For the Markov chain analysis we
 used $k=1,2$. 

In this case we found  a 7-word which appeared in the upstream
 region of all the seven genes: A fact that, looking at the probabilities listed
 in tab.3  certainly deserves more attention. The word is 
{\bf TTTAAAT}. A survey in the literature shows that this is indeed
 one of the binding sequences 
of a TF known as ``even-skipped'' which is known to regulate segmentation (and
also the development of certain neurons) in {\it Drosophila}. 
This TF has been widely
studied due to its crucial role in the early stages of embryo development, but
it was not directly related up to now to the regulation of glycolysis.
\end{description}
\begin{table}[t]
\caption{Probability $p(n,7)$
 of finding a $n$-word in the upstream region of all
the 7 genes involved in glycolysis. In the first column the value of $n$, in the
second the result obtained using the background probabilities. In the last two
columns the result obtained with the Markov chains with $k=1$ and $k=2$
respectively. 
\label{tab:three}}
\vspace{0.2cm}
\begin{center}
\begin{tabular}{|c|l|l|l|} \hline
$n$& $p(n,7)$ &  $p(n,7),~~ k=1$ &  $p(n,7),~~k=2$ \\ \hline\hline
6 & 0.346 & 0.76 & 0.78 \\ \hline
7 & 0.007 & 0.013 & 0.022 \\ \hline
8 & 0.00025 & 0.000034 & 0.00011 \\ \hline
\end{tabular}
\end{center}
\end{table}

\section{Example: diauxic shift in {\it S. cerevisiae}}
\label{s6}
As an example of the analysis of sect.\ref{s4}, we studied the 
so called {\it diauxic shift}, ({\it i.e.} the 
metabolic shift from fermentation to respiration),
in  {\it S. cerevisiae}
the pattern of gene expression 
during the shift was 
measured with DNA microarrays techniques in Ref. \cite{DeRisi:1997}.
In the experiment gene expression levels were measured for virtually
all the genes of at seven time-points while
glucose in the medium was progressively depleted.   As a result of our analysis
we found 
29 significant words, that can be grouped into 6 motifs ({\it i.e.}
groups of similar words).
Five of them correspond to known regulatory
motifs (for a database of known and putative TF's 
binding sites in S. cerevisiae see ref.~\cite{Pilpel:2001}).
 In particular three of them: STRE, MIG1 and UME6 (for the meaning of these
 abbreviations see again~\cite{Pilpel:2001}) were previously known to be
 involved in glucose-induced regulation process, 
while for the two other known motifs:
PAC and RRPE this was a new result. We consider the fact of having
found known regulatory motifs a strong validation of our method.

Finally we also found a new binding sequence:
{\bf ATAAGGG}, which we could not associate to any known regulatory motif.

\section{Conclusions}
\label{s7}

We have proposed two new methods to extract  biological information
 on the Transcription Factors (and more generally on the mutual interactions
 among genes) from
the statistical distribution of oligonucleotides in the upstream region of the
genes. Both are based on the notion of a ``regulatory network''
responsible for the various expression patterns of the genes, and aim
to find common binding sites for TFs in families of coregulated genes.

\begin{itemize}
\item
The methods can be applied both to selected sets of genes
of known biological
functions  (direct search
method) or to the genome wide {microarrays experiments}  (inverse search
method).
\item
They require a {complete knowledge of the upstream
 oligonucleotide sequences} and thus they
can be applied for the moment only to those organisms for which the complete
genoma has been sequenced.
\item
In the direct method, once the set of coregulated 
genes has been chosen, no further external input
is needed. The 
significance criterion of our candidates binding sites only depends
on the statistical distribution of oligonucleotides in the upstream region (or
in nearby regions used as test samples)
\item 
Both can be easily implemented and could be used as  standard preliminary
tests, to guide a more refined analysis.
\end{itemize}

Even if they already give interesting results, both our methods 
are far from being optimized. In particular there are three natural
 directions of improvement.
\begin{description}
\item {a]} Taking into account
the variability of the binding sequences,
\item {b]} Recognizing dyad like binding sequences (see for
 instance~\cite{vanHelden:2000}) which are rather common in eukaryotes,
\item{c]} Recognizing synergic interactions between TF's.
\end{description}
Work is in progress along these lines.
\vskip0.3cm
Needless to say the candidate binding sequences
 that we find with our method will have to be
tested experimentally. However our method could help
to greatly reduce the number of possible candidates and could
 be used as a guiding line for experiments.


\section*{References}
\begin{thebibliography}{99}

\bibitem{pg97}
M. Ptashne and A. Gann, {\it Nature} {\bf 386} (1997) 569

\bibitem{Wagner:1997} 
A. Wagner, 
{\it Nucleic Acids Research} {\bf 25} 3594-3604 (1997).

\bibitem{Tavazoie:1999}
S. Tavazoie, J.D. Hughes, M.J. Campbell, R.J. Cho and G.M. Church, 
{\it Nature Genetics} {\bf 22} 281-285 (1999).

\bibitem{Pilpel:2001}
Y. Pilpel, P. Sudarsanam and G.M. Church,
{\it Nature Genetics} {\bf 29} 153-159 (2001). Web supplement:\\*
http://genetics.med.harvard.edu/$\tilde{\ }$tpilpel/MotComb.html 

\bibitem{Bussemaker:2001} 
H.J. Bussemaker, H. Li and E.D. Siggia,
{\it Nature Genetics} {\bf 27} 167-171 (2001).


\bibitem{vanHelden:1998}
J. van Helden, B. Andr\'e and J. Collado-Vides, 
{\it J. Mol. Biol.} {\bf 281} 827-842 (1998).






\bibitem{vanHelden:2000}
J. van Helden, A. F. Rios and J. Collado-Vides,
{\it Nucleic Acids Research} {\bf 28} 1808-1818 (2000).

\bibitem{Hughes:2000} 
J. D. Hughes, P. W. Estep, S. Tavazoie and G. M. Church,
{\it J. Mol. Biol.} {\bf 296} 1205-1214 (2000).



\bibitem{hw2000}
R.Hu and B. Wang,
Archive: http://xxx.sissa.it/abs/physics/0009002



\bibitem{ccp} M. Caselle, F. DiCunto and P.Provero,
``Correlating overrepresented upstream motifs to gene expression: a
computational approach to regulatory element discovery in eukaryotes.''
Submitted to BMC Bioinformatics.


\bibitem{book}
B. Alberts  et al.,
{\sl Molecular Biology of the Cell }  (Garland Publishing Inc., New York, 
1994).


\bibitem{gag} 
R.C. Wilkins and J.T. Lis
{\it Nucleic Acids Research} {\bf 26} 2672-2678 (1998).


\bibitem{DeRisi:1997} 
J.L. DeRisi, V.R. Iyer and P.O. Brown, 
{\it Science} {\bf 278} 680-686 (1997).


\end{thebibliography}
\end{document}